\documentclass[10pt,preprint2]{aastex}
\received{2006 February~10}
\revised{2006 March~17}
\accepted{2006 March~21}
\paperid{20409}

\newcommand{\src}{\mbox{Cyg~A}}
\newcommand{\jybm}{\mbox{Jy~beam${}^{-1}$}}
\newcommand{\mjybm}{\mbox{mJy~beam${}^{-1}$}}

\shorttitle{Cygnus~A at Long Wavelengths}
\shortauthors{Lazio et al.}

\begin{document}

\title{Cygnus~A: A Long Wavelength Resolution of the Hot Spots}

\author{T.~Joseph~W.~Lazio, Aaron~S.~Cohen, Namir~E.~Kassim}
\affil{Naval Research Laboratory, 4555 Overlook Ave.~\hbox{SW}, Washington, DC
	20375-5351}
\email{Joseph.Lazio@nrl.navy.mil,Aaron.Cohen@nrl.navy.mil,Namir.Kassim@nrl.navy.mil}

\author{R.~A.~Perley}
\affil{National Radio Astronomy Observatory, \hbox{P}.\hbox{O}.~Box~0, Socorro, NM  87801}
\email{rperley@nrao.edu}

\author{W.~C.~Erickson}
\affil{Bruny Island Radio Spectrometer, 42~Lighthouse Rd., Lunawanna,
	Tasmania 7150 and School of Mathematics and Physics, University of Tasmania, Hobart, Tasmania 7005, Australia}

\author{C.~L.~Carilli}
\affil{National Radio Astronomy Observatory, \hbox{P}.\hbox{O}.~Box~0, Socorro, NM  87801}
\email{ccarilli@nrao.edu}

\and 

\author{P.~C.~Crane}
\affil{Naval Research Laboratory, 4555 Overlook Ave.~\hbox{SW}, Washington, DC
	20375-5351}
\email{Patrick.Crane@nrl.navy.mil}

\begin{abstract}
\noindent
This paper presents observations of Cygnus~A at~74 and~327~MHz at
angular resolutions of approximately 10\arcsec\ and~3\arcsec,
respectively.  These observations are among the highest angular
resolutions obtained below~1000~MHz for this object.  While the
angular resolution at~74~MHz is not sufficient to separate clearly the
hot spots from the lobes, guided by 151 and~327~MHz images, we have
estimated the 74~MHz emission from the hot spots.  We confirm that the
emission from both the western and eastern hot spots flattens at low
frequencies and that there is a spectral asymmetry between the two.
For the eastern hot spot, a low-energy cutoff in the electron energy
spectrum appears to explain the flattening, which implies a cutoff
Lorentz factor $\gamma_{\mathrm{min}} \approx 300$, though we cannot
exclude the possibility that there might be a moderate level of
free-free absorption.  For the western hot spot, the current
observations are not sufficient to distinguish between a free-free
absorped power-law spectrum and a synchrotron self-absorbed spectrum.
\end{abstract}

\keywords{galaxies: individual (Cygnus~A) --- radio continuum: galaxies}

\section{Introduction}\label{sec:intro}

Cygnus~A is the exemplar of a Faranoff-Riley Type~II radio galaxy,
marked by two large, limb-brightened lobes located nearly
symmetrically around a central core source.  Its evolution
\citep{br74,s74} is powered a ``central engine'' that ejects two,
oppositely directed relativistic jets, which are shocked and
terminated in ``hot spots''; driven by the momentum of the jets, the
hot spots move into the ambient medium, while the shocked material
flows out and ``behind'' the hot spots forming ``lobes.''  As one of
the nearest powerful radio galaxies, \src\ ($z = 0.057$,
\citealt*{ss82}) presents a valuable observational laboratory with
which to probe, at high resolution, the acceleration and emission
mechanisms within radio galaxies \citep{cb96,hc96}.

Using observations at frequencies from~151~MHz to~22~GHz,
\cite{cpdl91} performed a comprehensive analysis of the emission of
\src.  They confirmed that the lobes are powered by synchrotron
emission from particles accelerated in the hot spots and then
backflowing into the lobes.  Consistent with the notion that the hot
spots are burrowing their way into the ambient medium, the spectral
analysis by \cite{cpdl91} showed that the the emitting particles
nearer the core have a greater synchrotron age than those near the hot
spots.  They also showed that, above~1~GHz, the spectra of the hot
spots are consistent with relativistic particles being injected
continuously through a strong, nonrelativistic shock by diffusive
shock acceleration.

Below~1~GHz, \cite{cpdl91} found that the spectra of the hot spots
flatten.  They considered two explanations for this low-frequency
flattening, a low-energy cutoff (LEC) in the electron energy power
spectrum or synchtrotron self-absorption (SSA).  They favored the
former, both on observational grounds---model fits produced a superior
$\chi^2$ for LEC vs.\ SSA---and on theoretical grounds---the SSA model
implied magnetic field strengths within the hot spots of at least
3~\hbox{G}, orders of magnitude above the minimum-energy field
strengths.  \cite{cpdl91} found that an LEC implies a minimum electron
Lorentz factor of $\gamma_{\mathrm{min}} \approx 430$ in the hot
spots, assuming minimum-energy magnetic field strengths.
\cite{cpdl91} also considered and rejected thermal absorption by
foreground gas as an explanation for the spectral flattening as the
implied electron densities are far too high for internal absorption
and there is no optical H$\alpha$ emission from the cluster gas within
which \src\ is embedded.

The conclusion of an electron LEC within the hot spots relies
essentially on just two data, measured intensities at~151 and 327~MHz.
\cite{kpche96} conducted fairly low resolution observations
(25\arcsec) at~74~MHz and found a marked asymmetry, with the western
lobe being significantly weaker than the eastern lobe.  They
re-analyzed optical observations by \cite{cdcp89}, finding that the
amount of extinction in this direction may have been underestimated.
Thus, \cite{kpche96} attributed the asymmetry in the lobe emission to
thermal (free-free) absorption by foreground Galactic material,
postulated to be associated with a known H$\alpha$ filament.  However,
their observations did not have sufficient angular resolution to
separate the emission from the lobes and hot spots, so they were
unable to address the extent to which thermal absorption affects the
low-frequency spectra of the hot spots.

This paper presents new, higher-resolution 74 and~327~MHz observations
of \src\ made using the VLA in combination with the innermost antenna
of the Very Long Baseline Array (VLBA) connected to the VLA via
optical fiber link. The so-called ``VLA $+$ Pie Town link'' connected
interferometer provides an approximate factor of two improvement in
angular resolution over the original observations in \cite{cpdl91} and
\cite{kpche96}.  We use these higher resolution observation to revisit
the low-frequency spectra of the hot spots.


\section{Observations}\label{sec:observe}

Our observations were conducted on~2003 August~12 with the VLA in its
most extended configuration (A configuration) and with the VLBA Pie
Town antenna connected via a real-time optical fiber link \cite[the
``PT link,''][]{b00}. The longest baseline in the resulting array is
73~km.  Our observations were conducted simultaneously at~74
and~327~MHz.  Table~\ref{tab:summary} summarizes the observing and
image parameters.

\begin{deluxetable}{lcc}                                                
\tablecaption{Summary of Observing and Image
	Parameters\label{tab:summary}}
\tablewidth{0pc}
\tablehead{ 
\colhead{Observation:} & \colhead{74~MHz}  & \colhead{327~MHz}} 
\startdata                                   
Central Frequency (MHz) & 73.8            & 327.5 \\                            
Bandwidth (MHz)         & 1.5             & 3 \\
Channels (\#)           & 63              & 31 \\
Time on Source (hrs.)   & 13.6            & 13.6 \\                                
Resolution (\arcsec)    & $10.7\times7.1$ & $2.5\times2.5$ \\                
RMS noise (\mjybm)      & 970             & 24 \\
Peak Flux Density (\jybm) & 1040          & 113 \\                             
Image Dynamic Range     & 1070            & 4700 \\
\enddata                                                                
\end{deluxetable}                                                     

Spectral-line mode was used for both frequencies with~63 channels
across a 1.5~MHz bandwidth at~74~MHz and~31 channels across a 3~MHz
bandwidth at~327~MHz.  Spectral-line mode is the standard mode at both
frequencies because it is generally necessary to reduce bandwidth
smearing across the wide fields of view and to allow the removal of
radio frequency interference (RFI).
 
For both frequencies, existing models of \src, formed from previous
A-configuration observations, were used to calibrate the bandpass
response and complex antenna gains.  The flux densities of these
models are fixed to that found by interpolating the spectrum of
\cite{bgp-tw77}.  However, the distances from the Pie Town antenna to
the VLA antennas are from about~45 to~73~km, while the existing models
accurately represent the structure of \src\ only on angular scales
corresponding to baselines of~35~km or less, the size of the VLA A
configuration.  Consequently, we calibrated initially only with a
single short ($\approx 2$~min.) time interval near the end of the
observation when the elevation of \src\ was low enough that all
\emph{projected} baselines to Pie Town were shorter than 35~km;
the models were then able to accurately constrain that antenna's gain
and bandpass.  Models at higher resolution from higher frequencies did
not result in acceptable calibration, indicating that there are
significant changes in structure from higher frequencies.
 
Once calibrated, initial images were made.  Usually, 74 and~327~MHz
VLA observations require wide-field imaging to account for the
non-coplanar nature of the array.  Extending the baseline lengths to
the 70~km PT link would exacerbate the non-coplanar problem.  However,
\src\ is so much stronger than all other sources in the field of view
that it is not necessary to utilize wide-field imaging as essentially
no other sources can be detected.  Many rounds of phase-only
self-calibration followed by imaging were conducted at both
frequencies until the final images were obtained; amplitude
self-calibration was not used because experience shows that the gain
amplitude stability is quite good.

\begin{figure}
\epsscale{0.50}
\rotatebox{-90}{\plotone{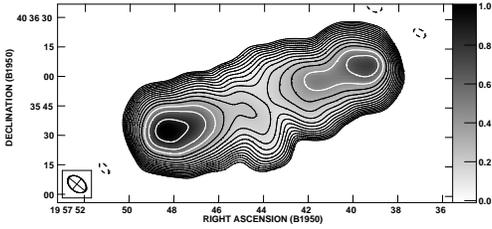}}
\vspace{-0.75cm}
\caption[]{\src\ at~74~MHz as observed with the VLA$+$PT link
interferometer.  The angular resolution is 10\farcs7 $\times$
7\farcs1, and the rms noise level is 0.97~\jybm.  The contours are
0.97~\jybm\ $\times -5$, 5, 7.07, 10, 14.1, \ldots, and the gray scale
is linear between~0 and~1000~\jybm.}
\label{fig:74}
\end{figure}

\begin{figure}
\rotatebox{-90}{\plotone{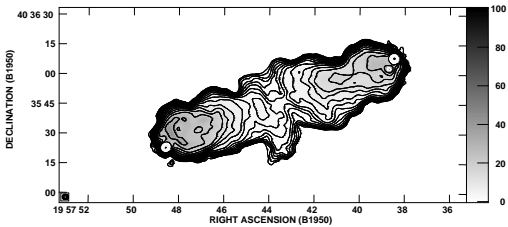}}
\vspace{-0.75cm}
\caption[]{\src\ at~327~MHz as observed with the VLA$+$PT-link
interferometer.  The angular resolution is $2\farcs5 \times 2\farcs5$,
and the rms noise level is 24~\mjybm.  The contours are 24~\mjybm\
$\times -10$, 10, 14.1, 20, 28.2, \ldots, and the gray scale is linear
between~0 and~100~\jybm.}
\label{fig:327}
\end{figure}

Figures~\ref{fig:74} and~\ref{fig:327} present our ``PT link'' images
at~74 and~327~MHz.  Because \src\ is so strong, these images are
dynamic range limited rather than thermal noise limited.

\section{Analysis}\label{sec:analyze}

We shall focus on the 74~MHz image for our analysis, as the flattening
of the hot spot spectra should be more pronounced in it. 
Even with the higher angular resolution provided by the PT link, the
resolution of the 74~MHz image is only 10\farcs7 $\times$
7\farcs1 and is insufficient to separate the hot spot and lobe
emission clearly.  In order to estimate the relative contributions of
the hot spots and lobes, we used a higher frequency image to estimate
a lobe contribution and remove this from the 74~MHz image.  This
method is similar to that used by \cite{cpdl91} to isolate the hot
spot emission at frequencies above~22~GHz in their analysis.

We formed a ``lobe-only'' image of \src\ at~151~MHz by modeling the
hot spots as gaussian components and subtracting them from a MERLIN
image with an angular resolution of~3\arcsec\ \citep{lms89}.  The
western hot spot could be represented adequately by two gaussian
components and the eastern hot spot by a single gaussian component.
We then convolved this lobe-only image to the same angular resolution
as our 74~MHz image and, assuming a spectral index, scaled it to
74~MHz.  Finally, we formed a residual image by subtracting the
frequency-scaled, lobe-only 151~MHz image from our 74~MHz image.
\cite{cpdl91} found a spectral index of~$-0.7$ for the lobe spectra
($S_\nu \propto \nu^\alpha$).  This spectral index appears to be too
steep, at least for the lobe emission near the hot spots, as it
produces systematically negative brightnesses for the lobe emission in
the residual image.  Requiring that the lobe emission in the residual
image be approximately 0~\jybm, we find a spectral index of~$-0.5$ is
more appropriate, which we adopted.  Overlaying the original 151~MHz
image on this residual 74~MHz image, we then selected only those
pixels at the locations of the hot spots from which to estimate the
hot spot emission at~74~MHz.

\cite{cpdl91} report the hot spot emission at higher frequencies as a
brightness (units of \jybm) for a circular 4\farcs5 beam, as this
resolution is sufficient to separate clearly the hot spot and lobe
emissions.  Because our 74~MHz beam is larger than 4\farcs5, we have
converted all brightnesses to units of Jy~arcsec${}^{-2}$.  In doing
so, we must also account for the possibility of beam dilution of the
hot spot emission.  Both by convolving the 151 and~327~MHz images to
the resolution of the 74~MHz image as well as by constructing a
synthetic image containing a 4\farcs5-diameter hot spot, we estimate
that the peak brightnesses in our 74~MHz image are diluted by a factor
of approximately 4.

Table~\ref{tab:fit} summarizes our estimates for the 74~MHz hot spot
emission, both the nominal values estimated from the image as well as
the dilution-corrected values.  In making these estimates, we analyzed
the two hot spots separately.  We base the uncertainties on our
analysis of the western hot spot.  We have modeled it with two gaussian
components, but a third component might also be used.
We repeated the above analysis for the western hot spot using three
gaussian components, and the difference between the analysis with two
or three components represents our estimated uncertainty.  For the
eastern hot spot, we have adopted the same uncertainty.

\begin{deluxetable}{lcc}
\tablecaption{\src\ 74~MHz Hot Spot Emission\label{tab:fit}}
\tablewidth{0pc}
\tablehead{
 \colhead{Hot Spot} & \colhead{$I_{\mathrm{peak}}$} &
	\colhead{corrected $I_{\mathrm{peak}}$} \\
                    & \colhead{(Jy~arcsec${}^{-2}$)} &
	\colhead{(Jy~arcsec${}^{-2}$)}}
\startdata                
East & 2.77 $\pm$ 0.58 & 11.08 $\pm$ 0.58 \\ 
West & 1.85 $\pm$ 0.58 &  7.40 $\pm$ 0.58 \\ 
\enddata
\tablecomments{$I_{\mathrm{peak}}$ is the nominal value estimated from
our 74~MHz image; the corrected value includes a factor of~4 increase
to account for beam dilution effects.}
\end{deluxetable}

\begin{figure}
\epsscale{0.55}
\begin{center}
\rotatebox{-90}{\plotone{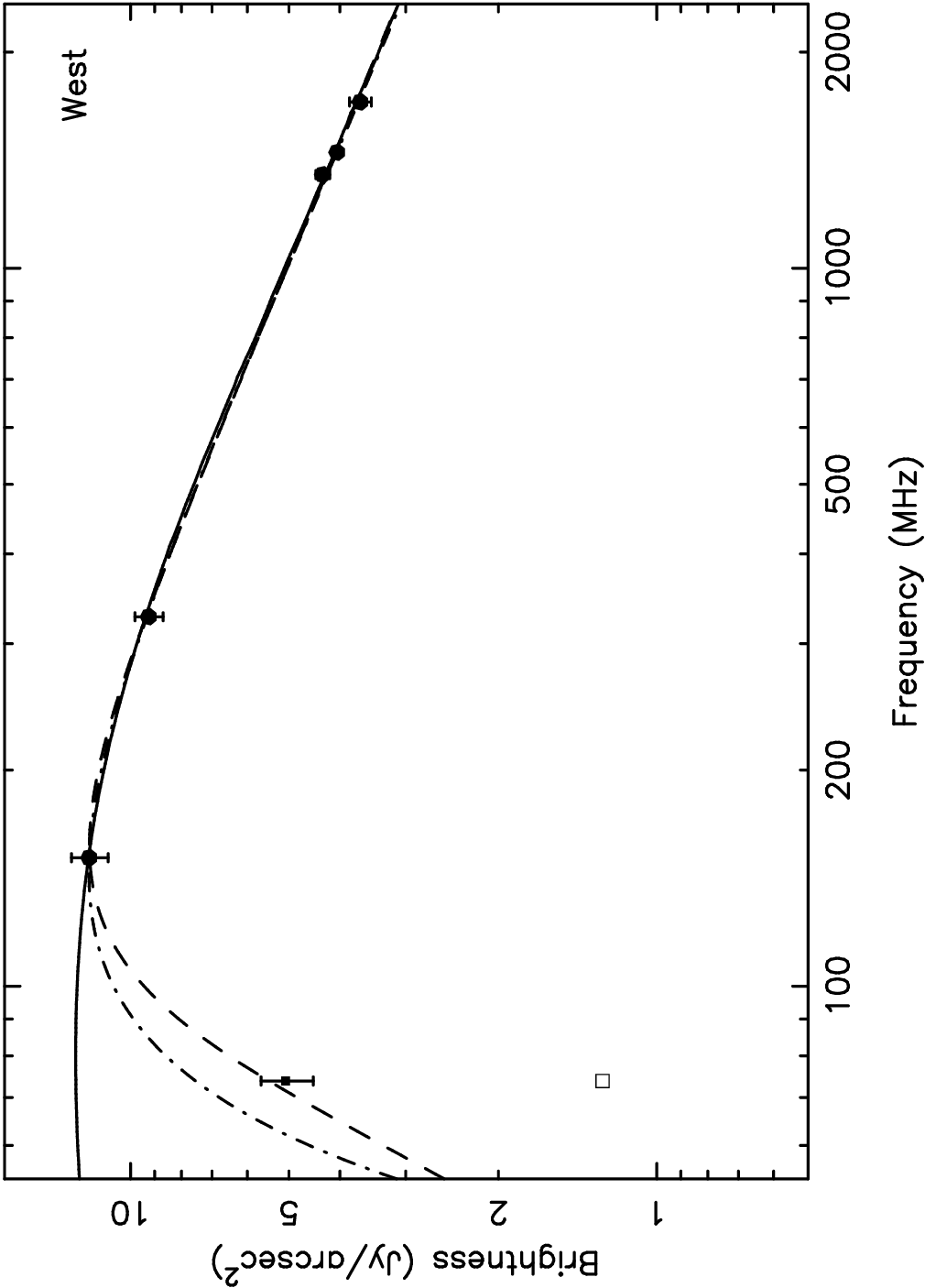}}\\
\rotatebox{-90}{\plotone{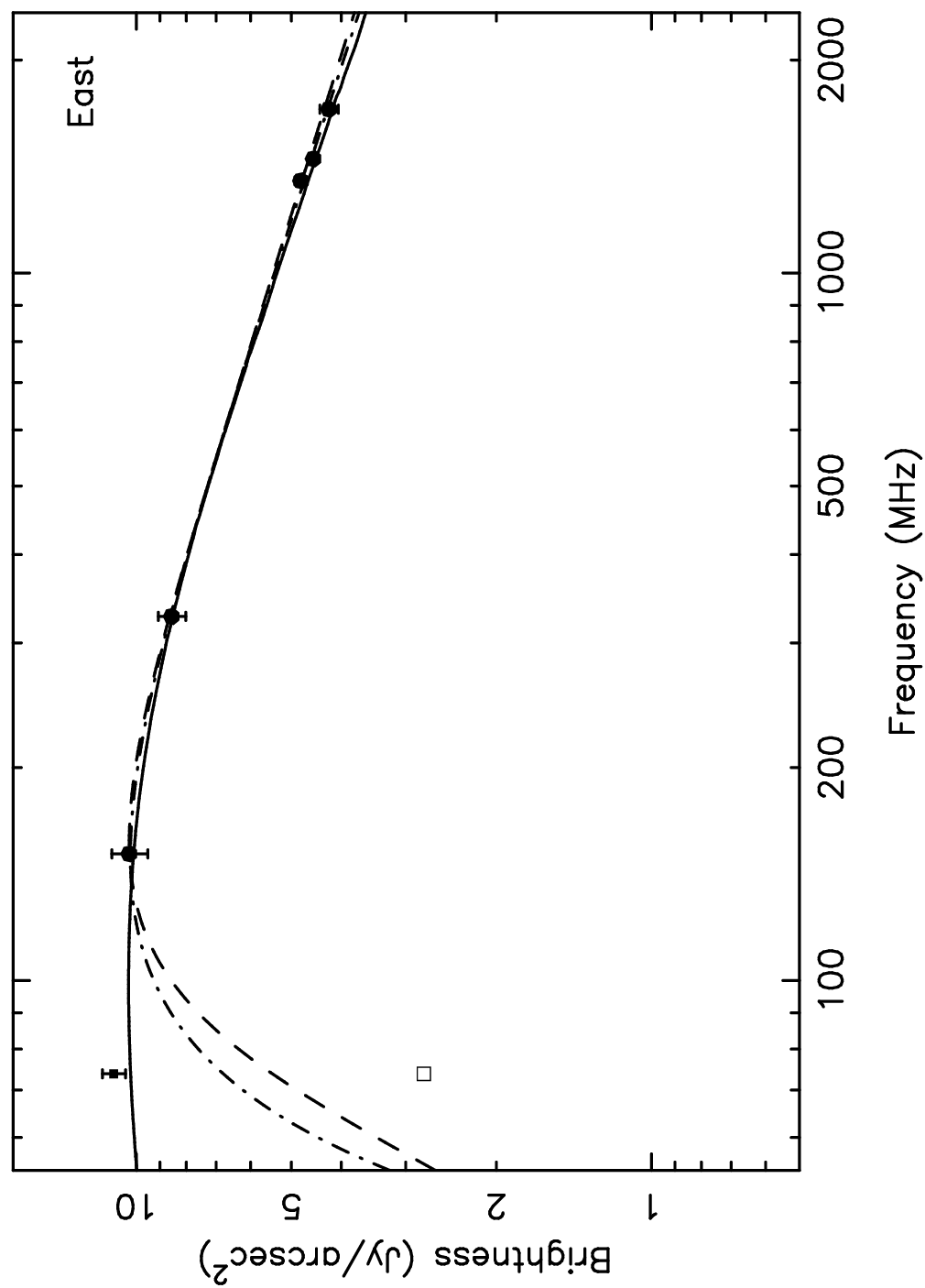}}
\end{center}
\vspace{-0.5cm}
\caption[]{The spectra of \objectname[]{Cyg~A}'s hot spots
below~2000~MHz.  Data at~151~MHz and above (solid circles) are from
\cite{cpdl91}.  The open square represents the apparent hot spot
brightness at~74~MHz, and the filled square is our estimate of the
brightness at~74~MHz accounting for beam dilution.
(\textit{Top}) The western hot spot.
(\textit{Bottom}) The eastern hot spot.
The solid curves show the expected frequency behavior for an electron
energy spectrum with a low-energy cutoff (LEC), the dashed curves
show the expected frequency behavior for synchrotron self-absorption (SSA),
and the dot-dashed curves show the expected frequency behavior for a
power-law spectrum with free-free absorption.
Table~\ref{tab:processes} lists the parameters for these fits.}
\label{fig:spectra}
\end{figure}

Figure~\ref{fig:spectra} shows the hot spot spectra below~2000~MHz.
Also shown are a spectrum with an \hbox{LEC}, an SSA spectrum, and a
free-free absorbed power-law spectrum.  Because the resolution
at~74~MHz does not yet approach that obtained at higher frequencies,
we have not included the 74~MHz data in producing the theoretical
curves.  In producing the LEC and SSA spectra, we have repeated the
analysis of \cite{cpdl91}, using formulae from \cite{p70}, but
restricted our attention to the data below~2000~MHz.  The free-free
absorbed power-law spectrum has been fit following the methodology of
\cite{k89}.  Table~\ref{tab:processes} summarizes the results of
fitting the higher frequency data for these different processes.

\begin{deluxetable}{lccc}
\tablecaption{Low Frequency Turnover Processes\label{tab:processes}}
\tablewidth{0pc}
\tablehead{
 \colhead{Process} & \colhead{Parameter} & \colhead{West hot spot} & \colhead{East hot spot}}
\startdata
LEC & $\alpha_{\mathrm{LEC}}$ & $-0.58$ & $-0.44$ \\
    & $\nu_{\mathrm{LEC}}$    & 110 $\pm$ 20~MHz & 120 $\pm$ 20~MHz \\
SSA & $\alpha_{\mathrm{SSA}}$ & $-0.57$ & $-0.43$ \\
    & $\nu_{\mathrm{SSA}}$    & 116~MHz & 106~MHz \\
free-free & $\alpha$          & $-0.59$ & $-0.45$ \\
          & $\tau_{74}$       & 1.1 $\pm$ 0.02 & 0.95 $\pm$ 0.02 \\
\enddata
\end{deluxetable}

\section{Discussion and Conclusions}\label{sec:conclude}

Immediately apparent is that the low-frequency spectra of both hot
spots flatten or turn over and that there is an asymmetry between the
two hot spots, with the eastern hot spot being approximately a factor
of~2 brighter than the western hot spot (\S\ref{sec:intro}).

For the eastern hot spot, both the SSA and the free-free absorbed
spectra considerably underpredict the 74~MHz (dilution-corrected)
brightness, while an LEC spectrum agrees reasonably well.  For an LEC
spectrum, we find a cutoff frequency of~120~MHz (Table~\ref{tab:fit}),
corresponding to a minimum Lorentz factor of $\gamma_{\mathrm{min}} =
320$.  This value is comparable to that found by \cite{cpdl91}, though
slightly lower because we have restricted our attention to lower
frequencies.  The free-free absorbed spectrum can be made to agree
with the 74~MHz datum by reducing the 74~MHz optical depth to
$\tau_{74} \approx 0.4$.  Clearly, however, an LEC spectrum can
explain the 74~MHz datum without requiring any free-free absorption.

For the western hot spot, the SSA and free-free absorbed spectra agree
reasonably well with the (dilution-corrected) 74~MHz datum, while
the LEC spectrum overpredicts it by a considerable amount.  On the
basis of the current data, we do not believe that it is possible to
distinguish between a free-free absorbed and SSA spectrum.  As
\cite{cpdl91} noted, the SSA spectrum  appears to require an  implausibly large magnetic field.  However,
this assumes a homogeneous emitting region in the source.  If the
emitting region is inhomogeneous, lower magnetic field values might be
tolerated at the expense of making the SSA turnover less sharp, and so
agree somewhat less well with the 74~MHz brightness.  Conversely, the
inferred optical depth of unity (Table~\ref{tab:processes}) requires
an emission measure of $\mathrm{EM} \sim 10^4$~pc~cm${}^{-3}$, well in
excess of the EM values observed \citep{cdcp89}.  \cite{kpche96}
found, based on lower resolution observations, that an emission
measure $\mathrm{EM} \sim 2500$~pc~cm${}^{-3}$ may be required to
explain the low-frequency spectrum of the western lobe, and there
could be clumps with larger values of EM across the face of the lobe,
though these would probably also mean that the reddening in this
direction has been underestimated substantially.  Finally, it may also
be possible that a free-free absorbed LEC spectrum, as opposed to a
free-free absorbed power-law as we have been considering previously,
could explain the low-frequency data.


Distinguishing between the various possibilities for the western hot
spot and confirming the apparent LEC in the spectrum of the eastern
hot spot will require additional observations.  More sensitive, high
resolution H$\alpha$ observations would constrain the amount of
foreground \ion{H}{2} while additional low radio frequency
observations at higher angular resolution would discriminate clearly
between the hot spot and lobe emissions.  At a minimum, the objective
should be to obtain angular resolutions comparable to the MERLIN image
($\approx 3\arcsec$), which will require baselines of roughly 200~km.
Multi-frequency observations, with a denser sampling of frequencies,
would be useful to constrain the effects of low energy cutoffs versus
free-free absorption, as well as probe the possibility of more complex
SSA models.  Future telescopes, notably the Long Wavelength Array
(LWA) and the Low Frequency Array (LOFAR), should provide much higher
resolution and a more densely sampled range of frequencies with which
to probe \src\ and similar objects.

\acknowledgements
We thank the many people at NRAO involved in making the
high-resolution capability of the Pie Town link a reality, P.~Leahy
for providing the 151~MHz MERLIN image, and D.~Harris for helpful
suggestions.  The NRAO is a facility
of the National Science Foundation operated under cooperative
agreement by Associated Universities, Inc.  Basic research in radio
astronomy at the NRL is supported by the Office of Naval Research.

\end{document}